\documentstyle[aps,psfig,multicol,eqsecnum,epsf,amssymb]{revtex}
\begin{document}
\draft
\title{Van der Waals loops and the melting transition in two dimensions}
\author{Juan J. Alonso} \address{
Departamento de F\'{\i}sica Aplicada I, Universidad de M\'alaga\\ 29071-M\'alaga, Spain\\}  
\author{Julio F. Fern\'andez}
\address{Instituto de Ciencia de Materiales de Arag\'on\\Consejo Superior de 
Investigaciones Cient\'{\i}ficas\\and Universidad de Zaragoza, 50009-Zaragoza, Spain\\} 
\date{\today}
\maketitle
\vspace{0.3cm} 
\begin{abstract}                
Evidence for the existence of van der Waals loops in pressure
$p$ versus volume $v$ plots has for some time supported the belief that
melting in two dimensions is a {\it first} order phase transition.
We report rather accurate equilibrium $p(v)$ curves for systems of hard
disks obtained from long Monte Carlo simulations.
These curves, obtained in the constant volume
ensemble, using periodic boundary conditions, exhibit
well defined van der Waals loops. We illustrate their existence
for finite systems that are known to undergo a {\it continuous}
transition in the thermodynamic limit. To this end, we obtain magnetization $m$ versus 
applied field curves from Monte Carlo simulations of the 2D Ising model, in the
constant $m$ ensemble, at the critical point. 
Whether van der Waals loops for disk systems behave in the $L\rightarrow\infty$
limit as they do for the 2D Ising model at the critical point cannot be
ruled out.
Thus, the often made claim that melting in 2D is a first order phase
transition, based on the evidence that van der Waals loops
exist, is not sound.

\end{abstract}  
\begin{multicols}{2}
Unphysical looking ``loops'' in pressure versus volume curves have been coming out of
approximate calculations for nearly a century\cite {vdw}. These so called
van der Waals loops have also been showing up in computer simulations of melting for over
three decades\cite {alder,brou,evans,tox,str,yama}. As Mayer and Wood pointed out
\cite {mayer}, pressures that increase with volume, that would be ruled out by van Hove's
theorem for {\it macroscopic} systems \cite {hove}, are indeed to be expected when
simulating melting of {\it finite} systems. van der Waals loops that decrease as system sizes
increase have been observed in simulations \cite{tox,yama}. Their existence
has almost invariably been taken 
as evidence of a first order phase transition \cite{alder,brou,evans,str} (though not always
\cite{tox}) and this has contributed much to the often held belief that the solid-fluid
phase transition in two dimensions (2D) is first order\cite{chester2,weber}. 

The purpose of this paper is threefold: (1) to give examples of van der Waals
loops that do sometimes show up for finite systems that
undergo {\it continuous} phase transitions in the thermodynamic limit;
(2) to point out that since their size (defined
below) is exactly equal to the free energy barrier for nucleation of the other phase, it
follows that van der Waals loops are to be taken as signs of {\it first} order transitions
only if their size vanishes in the thermodinamic limit as the inverse of the linear
system size ($L$); (3) to report accurate data for van der Waals loops that we have
obtained for two dimensional systems of $256$ and $1024$ classical hard disks, in the
fixed volume ensemble, and to show that their size dependence is in very good agreement
with more extensive data that follow from simulations in the
constant pressure ensemble (also known as the $NpT$ ensemble)\cite{wood,nos}
that seem to point to a second order transition, rather than to a first order one.  

The pressure $p(v)$ exerted by a system with a given fixed volume per particle $v$ is
usually obtained from Monte Carlo (MC) or from Molecular Dynamics simulations carried
out at constant volume. In order to obtain $p(v)$
one makes use of expressions that are derived
from the virial theorem \cite{metro}, which in turn follow from
the relation
\begin{equation}
p(v)=-{{\partial f(v,T)}\over{\partial v}},
\label{one}
\end{equation}
where $T$ is
the temperature and $f$ is the Helmholtz free energy per particle.
We have performed long Monte Carlo simulations 
($1.2\times 10^8$ MC sweeps in each run, of which the first $0.3\times 10^8$ sweeps
are allowed for equilibration) in the canonical ensemble
for systems of $256$ and $1024$ hard disks. The results obtained are shown in Fig 1 (as
$\circ$ and $\Box$ for $N=256$ and $1024$, respectively). 

Throughout this paper, ``volume'' $v$
actually stands for the {\it area} of a two dimensional system; it is given in terms
of the closest packing area $v_0$, and has therefore no units.
The pressure $p$ is actually a force per unit length, which we give in terms
of $kT/v_0$ and has therefore no units.
 
There is an alternative
way to obtain the same function, $p(v)$, that illustrates how van der Waals loops
come about for finite systems. Consider the
probability density, $P_p(v)$, that a system at a given pressure $p$ have specific
volume $v$. $P_p(v)$ can be obtained through Monte Carlo simulations carried out at a
given pressure $p$, in the $NpT$ ensemble. Data for $P_p(v)$ that
have been obtained\cite{nos} for a system of $256$ hard disks in the
solid and fluid phases are shown in Fig. 2a. Data for $P_p(v)$ exhibiting
coexistence of both phases are shown in Fig. 2b \cite{nos}.
The $p(v)$ curve that ensues in the {\it canonical} ensemble (that is, $-\partial f/\partial
v$), can be obtained from $P_p(v)$, since $P_p(v)$ and $f(v)$ are related by
\begin{equation}
P_p(v) \propto \exp\{-[f(v)+pv]N/kT\},
\label{P_p}
\end{equation}
where $k$ is Boltzmann's constant. Each data point $(v,p)$ exhibited in Fig. 1 either
as a $\bullet$ or as a $\blacksquare$, and in Fig. 3 as a $\bullet$, has been obtained
from an independent MC run at the corresponding value of $p$.
Note that any two volumes, such as $v_a$ and $v_b$ in Fig. 3, that fulfill $p(v_a)=p(v_b)$
are most probable volumes (see Fig. 2) when $p$, instead of $v$, is fixed.
On the other hand, $v_c$, which is the portion of the loop where
$\partial p/\partial v>0$ and satisfies
$p(v_c)=p(v_a)=p(v_b)$, is the least probable volume. 

Alternatively, $f(v)$ may, of course, be extracted from
(at least in principle) $P_p(v)$ obtained from a single simulation
at an arbitrary constant $p$, using Eq. (2).
Pressure versus volume curves follow from Eq. (1). Data points thus obtained,
from $P_p(v)$ for $p=7.64$ and $N=256$ (exhibited in Fig. 2b),
are shown in Fig. 3 (as $\Box$). The good agreement between
independent sets of the data points in Figs. 1 and 3
gives an indication of the accuracy of our equilibrium results.

For comparison, we also plot in Fig. 3 $p$ versus
the mean volume $\langle v\rangle$ that is obtained in the $NpT$ ensemble
for a system of $256$ disks (shown as $\Diamond$).
No van der Waals loops obtain. This is because, in the $NpT$
ensemble, $\partial \langle v\rangle /\partial p=-N\langle
(v-\langle v\rangle )^2\rangle /kT$, which is clearly negative.

We next give an example that underscores the fact that while van der Waals loops
follow for finite systems from first order phase transitions, the converse is {\it not} true.
Consider the Ising model
in 2D. The applied field $h$ versus magnetization, $m$, is shown in
Fig. 4 for Ising systems of $64, 256$ and $1024$ spins, for periodic boundary conditions,
at the critical temperature $T=T_c$. 
These curves are unusual because we obtained them from MC
simulations in an constant $m$ ensemble,
rather than in the more often used ensemble in which $h$ is fixed.
In our simulations, we keep $m$ constant using the Kawasaki algorithm \cite {kawa}.
In it, one spin
is flipped up while another spin is flipped down at each MC step.
The canonical and $NpT$ ensembles, discussed above, correspond to the constant
$m$ and constant $h$ ensembles, respectively.

We arrive at the $h(m)$ values for the constant magnetization
ensemble given in Fig. 4 as follows.
Consider first a simulation performed at constant $h$, in which
spins are flipped, and the total magnetization $M$ (given by $Nm$)
is consequently not conserved.
Let $w(m+2/N\leftarrow m)$ be the conditional probability that a
system known to have
an $M$ value make the transition $M+2\leftarrow M$ when no external field
is applied.
Since $P_h(m)\propto \exp [-Nf(m)/kT]$ for $h=0$, we can write the
detailed balance condition, 
\begin{equation}
{{w(m+2/N\leftarrow m)}\over
{w(m\leftarrow m+2/N)}}={{e^{-Nf(m+2/N)/kT}}\over{e^{-Nf(m)/kT}}}.
\end{equation}
Let the Hamiltonian ${\cal H}$ become ${\cal H}-hM$ when field $h$ is applied.
Now, $h$ plays no role in a calculation in the constant $M$ ensemble,
but an $h(m)$ can be obtained for the given value $m$ from
the relation $h=\partial f(m)/\partial m$.
Taking logarithms of both sides of the above equation
gives $f(m+2/N)-f(m)$ in terms of the transition
rates. The approximation $f(m+2/N)-f(m)=(2/N)[\partial f(m+1)/\partial m]$
gives 
$h(m+1)=(N/2)[f(m+2/N)-f(m)]$. We thus arrive at
\begin{equation}
h(m)=-{{kT}\over{2}}\ln\left[{{w(m+1/N\leftarrow m-1/N)}\over
{w(m-1/N\leftarrow m+1/N)}}\right],
\end{equation}
after shifting $m\rightarrow m-1$ for symmetry's sake.
In order to obtain the transition rates, we proceed as follows.
First note that the probability for an up spin flip from a given
spin configuration is proportional to
either 1 or $\exp(-\Delta E/kT)$, depending on whether the corresponding
energy change $\Delta E$ is either negative or positive, respectively.
Accordingly, after each MC sweep, having applied Kawasaki's rule throughout the
entire system, we assign to each spin down either the number $1$ or the number
$\exp(-\Delta E/kT)$, if flipping it up would lower its energy or
raise it by $\Delta E$, respectively. (No spin is actually flipped.)
The sum of such numbers [$1$ and $\exp(-\Delta E/kT)$] over all down spins
in the system averaged over an MC run is our unnormalized estimate
of $w(m+1/N\leftarrow m-1/N)$.

Alternatively, the same curves for $h(m)$ may be obtained from simulations
in which $h$ is fixed, by applying the relation $h=\partial f(m)/\partial m$
to probability curves $P_h(m)$ that follow from such simulations.
Results obtained in this fashion, from both the constant $m$ and the constant $h$
ensembles, are exhibited in Fig. 4.

We have also obtained $h(m)$ curves (not shown) for $T<T_c$ for the 2D Ising model.
However, loop sizes for $T<T_c$ and for $T=T_c$
vary rather differently with system size. By loop size, we
mean the free energy $\Delta g=\int h(m)dm$ over the domain defined by $m<0$ and $h>0$.
Data points for $L\Delta g/kT$ are shown in Fig. 5 for $T/T_c=0.9$, $0.95$ and $1$.   
Data points for the Gibbs free energy
\begin{equation}
\Delta g=\int_{v_a}^{v_c}[p(v_a)-p(v)]dv
\label{Dg}
\end{equation}
(the shaded area in Fig. \ref{Fig3}) are also shown in Fig. \ref{Fig5}
for disk systems of various sizes.
In order to make $\Delta g$ unique, we choose $v_a$ such that $p(v_a)$ is the
Maxwell construction pressure, $p_m$ \cite{maxc}, that is,
$\int_{v_a}^{v_b}[p(v)-p_m]dv=0$.
>From data points shown in Figs. 1 and 3, we obtain, making use of Eq. (3),
the two data points shown in Fig. 5 as $\bullet$ for systems of
$L\times L$ spins for $L=16$ and $L=32$.
Data from previous simulations in the $NpT$ ensemble
are also shown (as $\circ$) for $L=16$, $20$, $24$ and $32$ \cite{nos}.
The error
bars shown in Fig. 5 follow from the procedure described in the Appendix.

As Lee and Kosterlitz have explained in some detail \cite{koster},
the macroscopic limit $L\Delta g$ (a measure of the surface tension) does
not vanish for first order phase transitions. This follows from the following
simple argument. As may be seen from Eq. (\ref{Dg}) and comparison of
Figs. \ref{Fig2} and \ref{Fig3},
$L^d\Delta g$ (the system's dimension $d$ is $2$ in this case)
is the free energy barrier that is surmounted by the system
when a fluctuation takes it over the top, at $v_c$, starting
from the (locally) most probable volume $v_a$. We may therefore think of $L^d\Delta g$
as the free energy of the wall that arises between two coexisting phases when one
of them is nucleated from the other one in order to make a transition
from one phase to the other one. Since the wall
thickness is finite for first order transitions, it follows then that
$L^d\Delta g\sim L^{d-1}$, which is the desired result.

Inspection of Fig. 5 shows that
vanishing of $L\Delta g$ in the $L\rightarrow\infty$ limit,
as for the 2D Ising model at the critical point, can
clearly {\it not} be ruled out for disk systems. 
Thus, the often made claim that melting in 2D is a first order phase
transition, based on the evidence that van der Waals loops
exist\cite{alder,brou,evans,str}, is not
sound. Further results for larger systems would help to establish how the
$L\rightarrow\infty$ limit of the surface tension behaves.

It is perhaps worth stating explicitly that
whereas phase coexistence and nonvanishing surface tension that are associated
with first order transitions imply van der Waals loops, their appearance
depends on boundary conditions when continuous phase transitions are involved. 
Indeed, whereas $P_h(m)$ for the 2D Ising
model at the critical point is bimodal for periodic boundary conditions, it exhibits
one single maximum (no van der Waals loops then) for free boundary
conditions \cite{binder}. Analogously, no van der Waals loops are obtained
for systems of disks for hard crystalline walls\cite{nos} for $N\leq 4096$ 
or for systems of disks on spherical surfaces\cite{moore}. This provides
support for the proposition that melting of systems of hard disks 
in 2D unfolds through a continuous transition.
 
J. J. A. and J. F. F. are grateful for partial financial support from DGES of Spain,
through grants Nos. PB97-1080 and PB95-0797, respectively.
It is a pleasure to acknowledge continued help with computer work from
Dr. Pedro Mart\'{\i}nez, ample computer time from CECALCULA, at Universidad
de Los Andes.

\appendix
\section{}

We specify here how we obtain the error bars shown in Figs. 1 and 5.

The error bars shown in Fig. 1 for data that follows from MC runs in the
canonical ensemble are obtained as follows. We divide each MC run into
5 ``time'' intervals, and calculate an average pressure value for
each one of the five intervals. Twice the values of the standard
deviations thus obtained from each such set are exhibited in Fig. 1 as error bars.

There are two kinds of error bars shown in Fig. 5. We first discuss
the ones that follow from MC runs in the $NpT$ ensemble. We have obtained
the probabilities $P_p(v)$ from independent MC runs for various values of $p$.
As discussed in the text, the free energy $f(v)$ is given by,
\begin{equation}
e^ {-Nf(v)/kT}=P_p(v)e^{Npv/kT}.
\end{equation}
Using six curves for $f(v)/kT$ thus obtained,
we exhibit $N[f(v)+p_mv]/kT$ in Fig. 6, for various values of $p$ and $p_m=7.865$,
for systems of $1024$ disks. 
Slightly different values of $L\Delta g/kT$ are obtained from
each one of those $f(v)+p_mv$ curves (see Fig. 6), from which the standard deviation is obtained.
It is shown as the error for $L=32$ in Fig. 5.
Other error bars shown in Fig. 5 for the values of $L\Delta g/kT$ that follow from
the $NpT$ ensemble are obtained similarly.

We next explain how we obtain the error bars for the
data points for $L\Delta g/kT$ shown as $\bullet$ in Fig. 5. 
These errors follow from the errors shown in Fig. 1 for values of
$p(v)$ that were obtained from simulations in the {\it canonical} ensemble. An explanation is
called for because the error evaluation procedure is not trivial:
variations in $p(v)$ lead to variations in the Maxwell
construction pressure and in the integration limits of Eq. (\ref{Dg}).
However, a bit of reflection shows that such changes in the limits of integration
give contributions to errors in $\Delta g$ that are of second order in $\delta p(v)$
[where $\delta p(v)$ is the difference between an erroneous pressure and the
correct one]. 
Accordingly, we shall neglect variations in $v_a$, $v_c$, and $v_b$
(which are defined as in Fig. 1). Thus, the first order change, $\delta p_m$,
that is induced in the Maxwell construction pressure $p_m$, by errors in $p(v)$,
is given by, 
\begin{equation}
(p_m+\delta p_m)(v_b-v_a)=\int_{v_a}^{v_b}[p(v)+\delta p(v)]dv,
\end{equation}
which, by virtue of the definition of $p_m$ itself, leads to,  
\begin{equation}
\delta p_m={1\over{(v_b-v_a)}}\int_{v_a}^{v_b}\delta p(v)dv.
\end{equation}
Now, the first order variation $\delta \Delta g$, which follows from
Eq. (\ref{Dg}), is given by,
\begin{equation}
\delta (\Delta g)=\delta p_m(v_c-v_a)-\int_{v_a}^{v_c}\delta p(v)dv,
\end{equation}
which, upon substitution of $p_m$ from Eq. (A3), becomes
\begin{equation}
\delta (\Delta g)={1\over{(v_b-v_a)}}[(v_c-v_a)
\int_{v_c}^{v_b}\delta p(v)dv-(v_b-v_c)
\int_{v_a}^{v_c}\delta p(v)dv].
\end{equation}
Finally, we replace integrals by sums. Furthermore, we note that all errors
obtained for $p(v)$ for different values of $v$ follow from different MC runs
and are therefore assumed to be statistically independent. We thus arrive at
the average value of $[\delta(\Delta g)]^2$, 
\begin{equation}
\langle [\delta (\Delta g)]^2\rangle^{1/2}={{\Delta v}\over{(v_b-v_a)}}(q+s),
\end{equation}
where
\begin{equation}
q=(v_c-v_a)
[\sum_{v_c<v_i<v_b}\delta p(v_i)^2]^{1/2},
\end{equation}
\begin{equation}
s=(v_b-v_c)[\sum_{v_a<v_i<v_c}\delta p(v_i)^2]^{1/2}\},
\end{equation} 
and $\Delta v=v_{i+1}-v_i$. This is the desired expression.

\end{multicols}

\newpage
\begin{figure}
\centerline{\psfig{figure=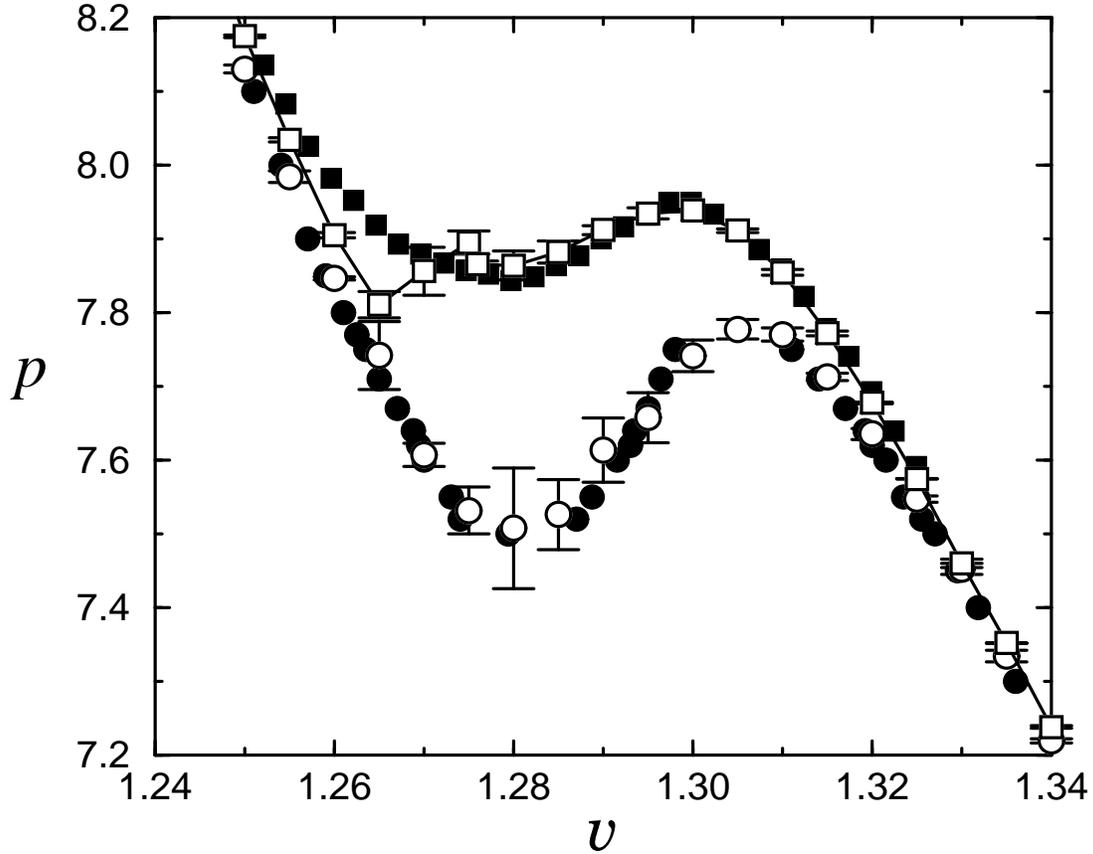,width=16cm,angle=0}}
\caption{Pressure $p$ versus volume $v$ data points from MC simulations
systems of $N$ disks. The ``volume'' $v$ stands for an {\it area}, it is given in terms
of the closest packing area $v_0$. The pressure $p$ is a force per unit length,
given in terms of $kT/v_0$. Neither $v$ nor $p$ have therefore any units.
$\circ$ and $\Box$ stand for results from
simulations in the constant volume ensemble for $N=256$ and $N=1024$, respectively.
$\bullet$ and $\blacksquare$ stand for results, extracted from
simulations in the $NpT$ ensemble {\it making use of Eqs. (1) and (2)},
for $N=256$ and $N=1024$, respectively.
All data points follow from runs of approximately $10^8$ MC sweeps, after
equilibrating the system for $3\times 10^7$ MC sweeps. For details about the
error bars shown, see the Appendix.}
\label{Fig1}
\end{figure}
\newpage
\begin{figure}
\centerline{\psfig{figure=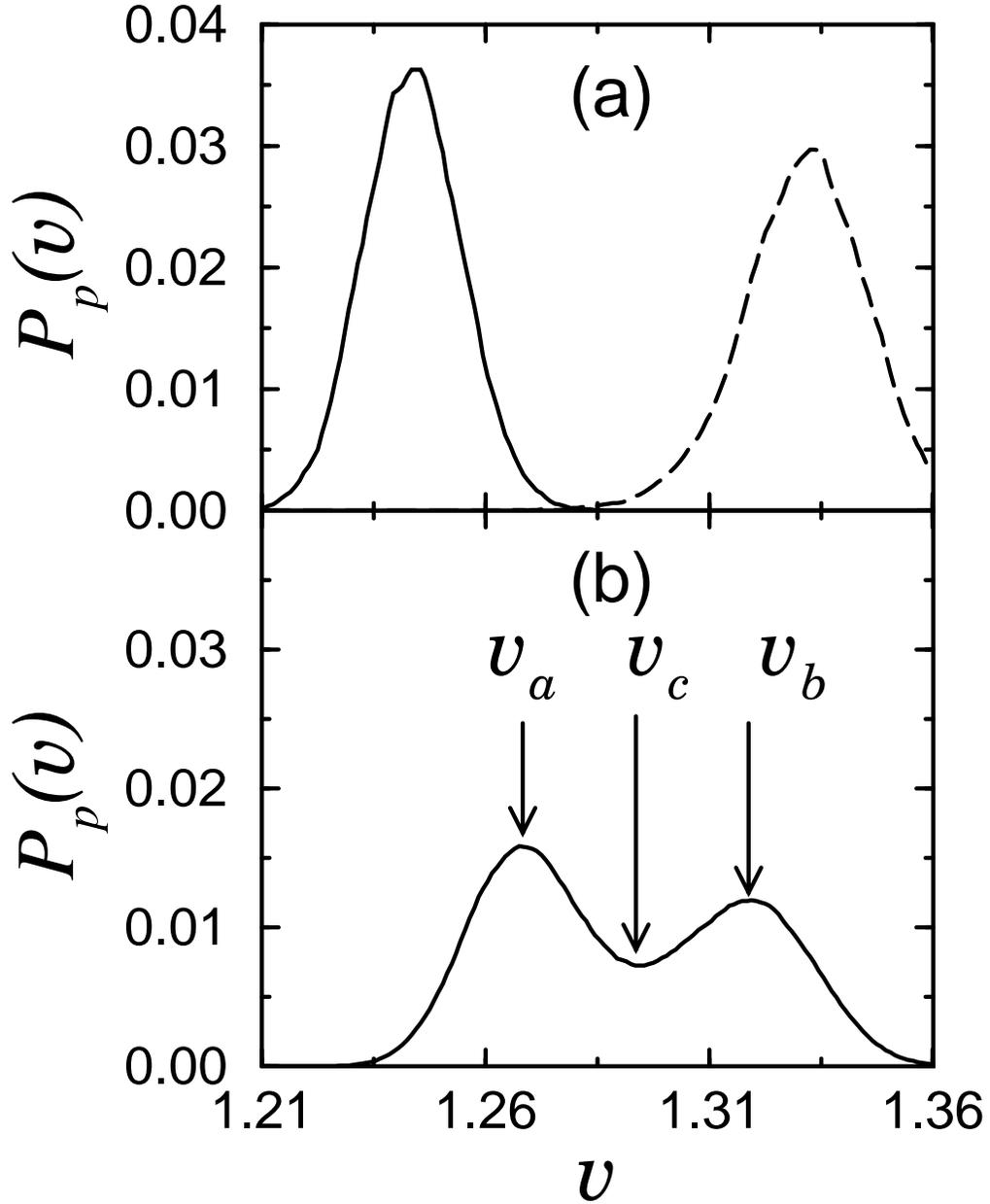,width=16cm,angle=0}}
\caption{(a) Frecuency of occurrence $P_p(v)$ for specific volume $v$
for a system of $N=256$ for $p=7.83$ (solid line) 
and for  $p=7.40$ (dashed line). The units for $v$ and $p$ are given in the
caption for Fig. 1. (b) Same as for (a) but for
$p=7.64$. These curves follow from runs of over
$2\times 10^8$ MC sweeps. Lines shown go through datapoints obtained, one for each
$\Delta v =10^{-3}$ bin. Volume values where
$\partial [f(v)+pv]/\partial v=0$ are marked with arrows.}
\label{Fig2}
\end{figure}
\newpage
\begin{figure}
\centerline{\psfig{figure=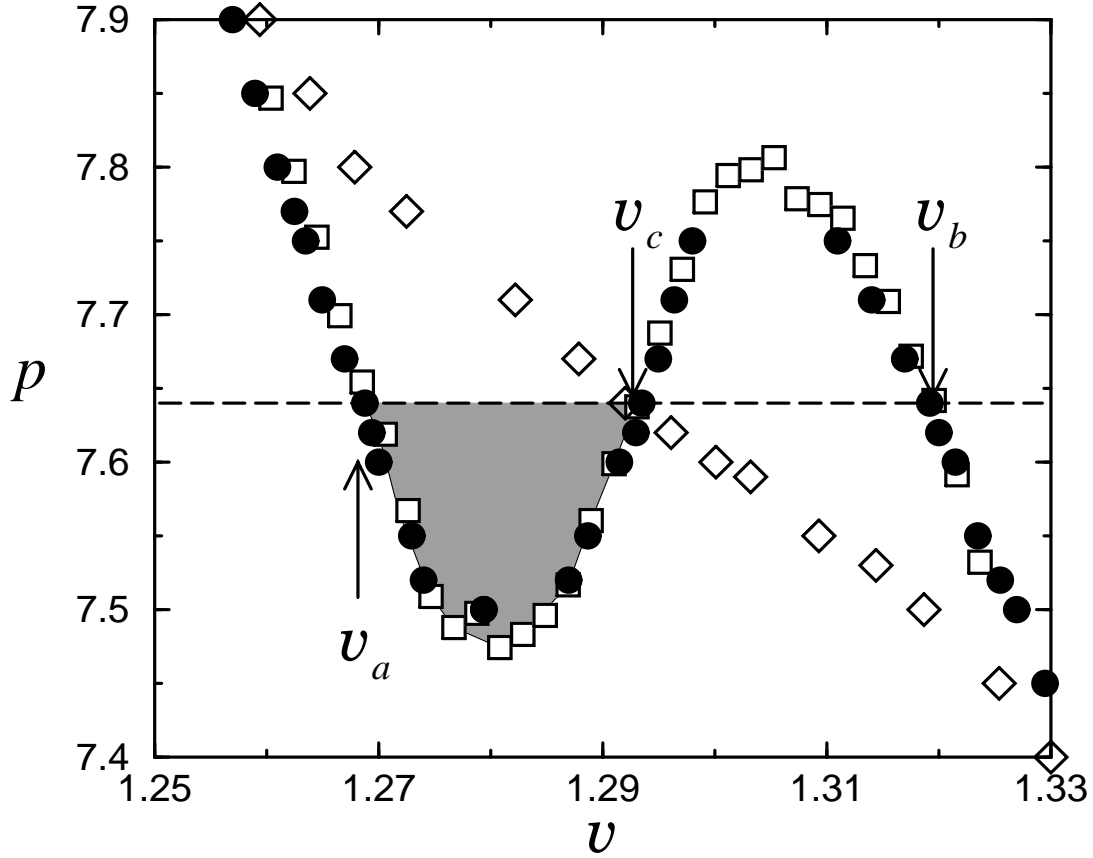,width=16cm,angle=0}}
\caption{Data points for $p$ vs volume for systems of $N=256$ disks.
$\Diamond$ stand for data points $(<v>,p)$ obtained from
simulations using the $NpT$ ensemble. The units for $v$ and $p$ are given in the
caption for Fig. 1.
$\Box$ are for the numerically obtained derivative $- \partial f(v) / \partial v$ from
the frecuency of occurrence $P_p(v)$ for a single pressure, $p=7.64$, value.
$\bullet$ are for points $(p_0,v)$ fulfilling the relation
$\partial [f(v)+p_0 v]/\partial v=0$
where $f(v)+p_0 v$ follows from
$\ln [P_p(v)]$ for $p=p_0$. For example, for $p=7.64$ (marked with
a dashed line in the figure) we find three different solutions $(v_a,v_c,v_b)$
(marked with arrows as in fig. 2b).}
\label{Fig3}
\end{figure}
\newpage
\begin{figure}
\centerline{\psfig{figure=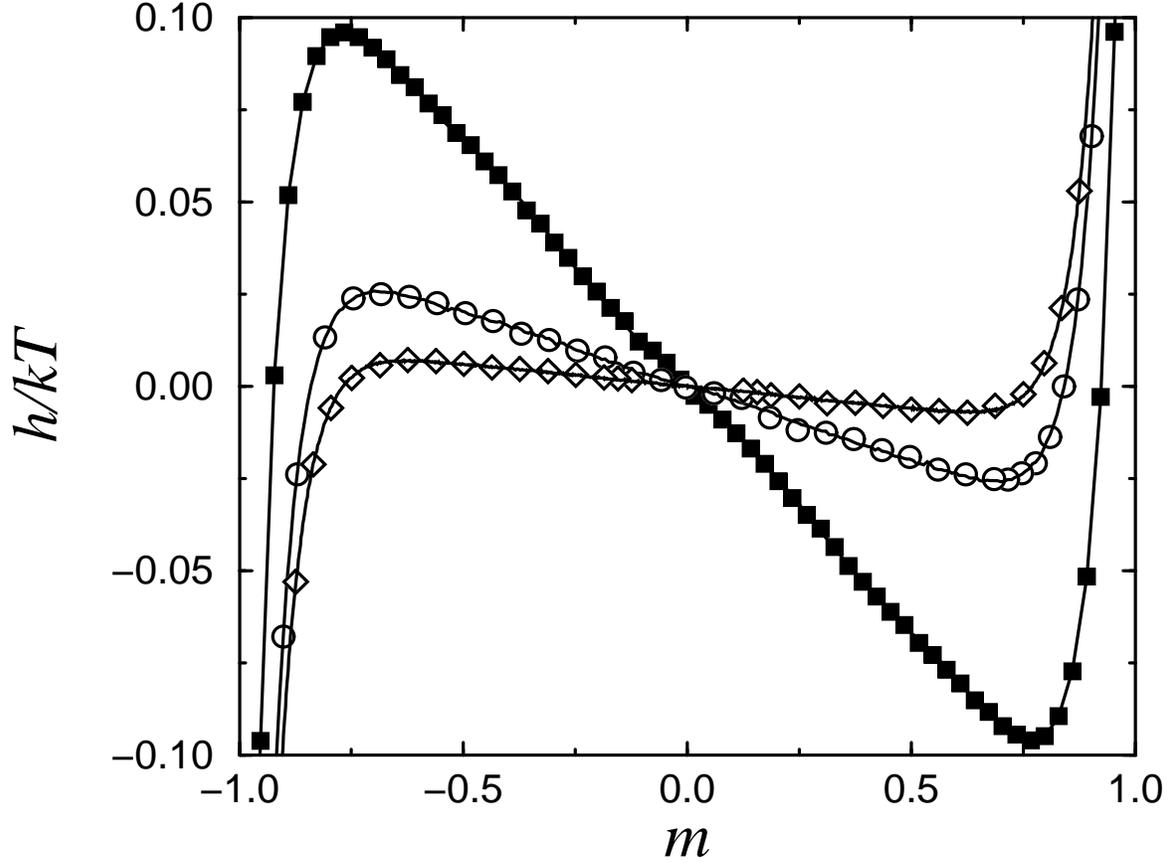,width=16cm,angle=0}}
\caption{Magnetic field $h$
versus magnetization $m$ for the $2D$ Ising systems of $L\times L$ spins at the critical
temperature, for $L=8, 16$ and $32$. Continuous
lines stand for data that follows from probability $P_h(m)$ curves, 
obtained from simulations in the constant $h$
ensemble for $h=0$. The umbrella method was used to obtain $P_h(m)$, covering
the whole range of $-1\leq m\leq 1$ values with $16$ ``umbrellas''.
Each one of the three continuous lines shown follows from $16$ MC runs
of $10^8$ sweeps over the entire system for $N\geq 12$ and $5\times 10^6$ sweeps
for $L=8$. 
$\blacksquare$, $\circ$, and $\Diamond$ stand for $L=8, 16$ and $32$ respectively,
obtained in the conserved magnetization $m$ ensemble (with the Kawasaki algorithm).
Every point shown follows from a MC run of $5\times 10^6$ sweeps for $L=8$,
and $10^8$ sweeps for $N\geq 12$.}
\label{Fig4}
\end{figure}
\newpage
\begin{figure}
\centerline{\psfig{figure=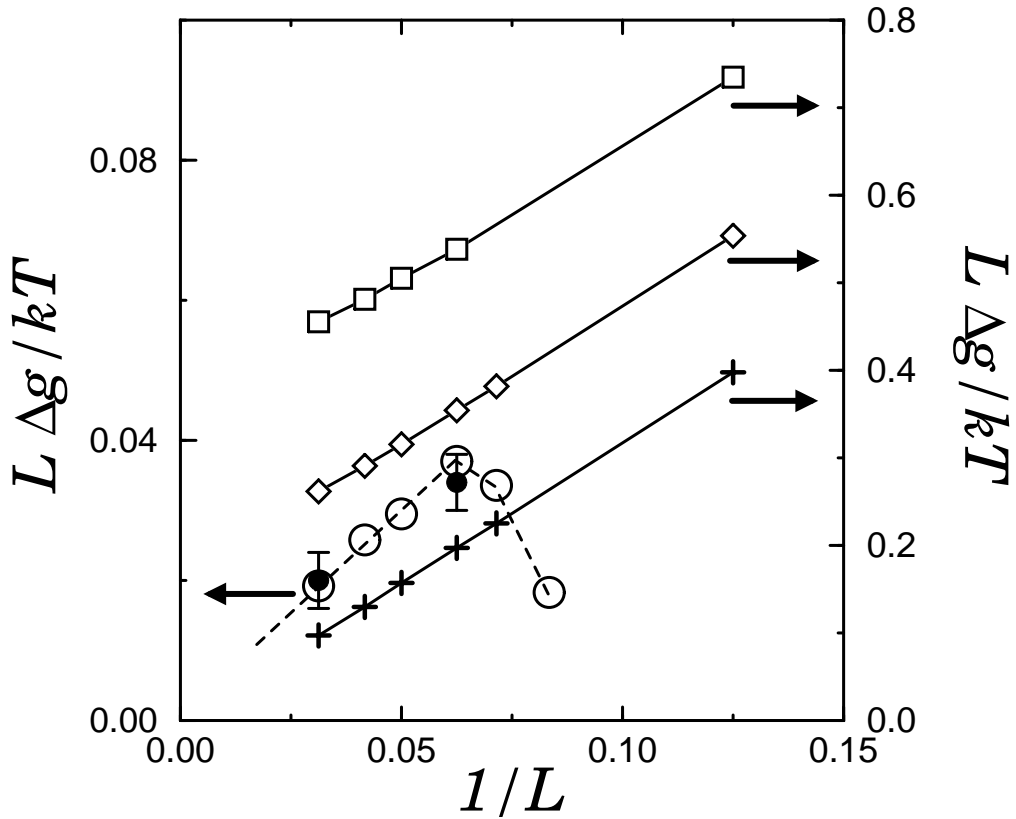,width=14cm,angle=0}}
\caption{Data points for $L\Delta g/kT$ for a system of $L\times L$ disks.
$\bullet$ and $\bigcirc$ stand for data obtained from systems of $L\times L$ hard disks
in the constant volume and $NpT$ ensembles, respectively.
$\Diamond$, $\Box$ and $+$
stand for the 2D Ising model at temperatures $T/T_c=0.9, 0.95$, $1$,
respectively. Lines are only guides to the eye.
Error bars for data points shown as $\bigcirc$ are approximately given
by the size of the circles, except that the error is ($\pm 0.0006$)
approximately 4 times smaller for $L=32$.
For a detailed account about how these errors,
as well as the error bars shown for the $\bullet$ data points, were
obtained, see the Appendix.}
\label{Fig5}
\end{figure}
\newpage
\begin{figure}
\centerline{\psfig{figure=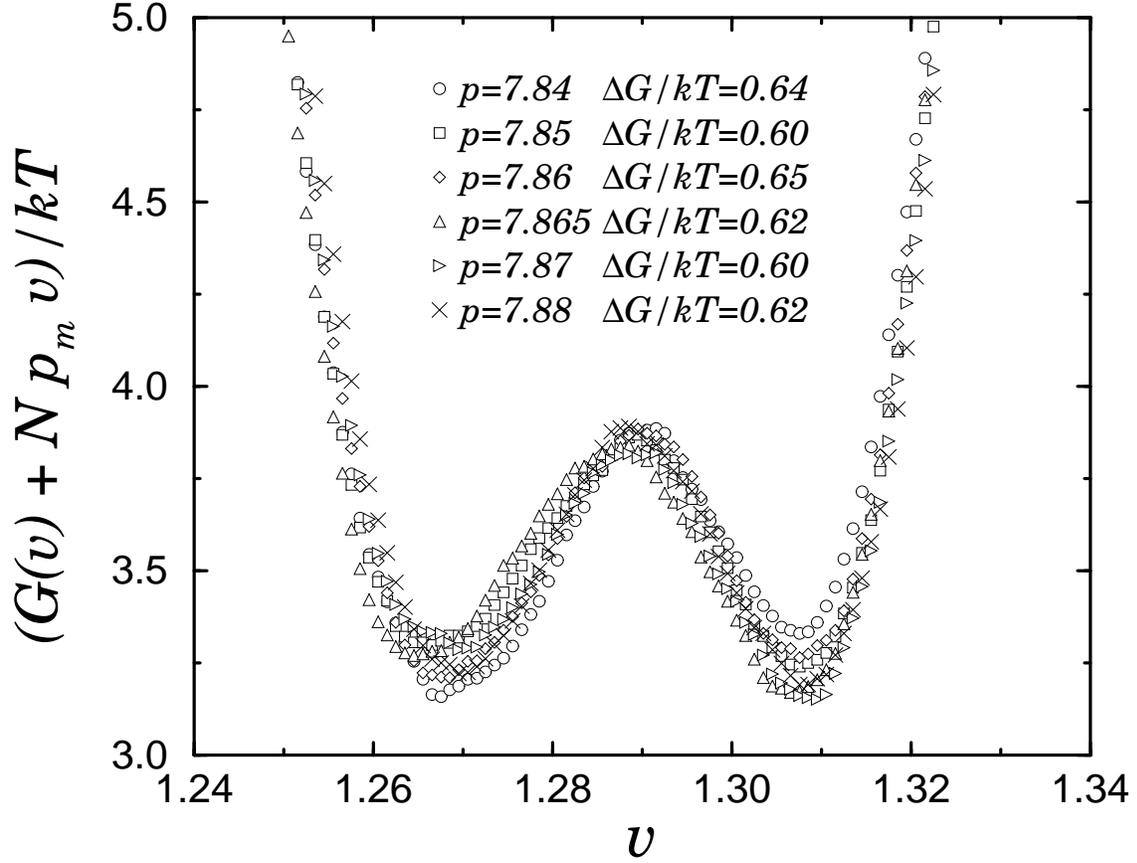,width=16cm,angle=0}}
\caption{Quantity $N[f(v)+p_mv]/kT$ versus $v$, up to a constant, obtained from
MC simulations of systems of $1024$ disks for each of the values of $p$ shown;
$p_m=7.865$. Each one of the six curves shown follows from a MC run of at least
$2\times 10^8$ MC sweeps over the whole system. Ideally, all curves should be
equal. We define the free energy barrier,
$\Delta G$ as the local maxima of $N[f(v)+p_mv]$ minus the average value of its
two minimum values. All values of $\Delta G/kT$ thus obtained are shown in the figure.
We obtain the average value $\Delta G/kT=0.62$. The corresponding standard
deviation is 0.02.}
\label{Fig6}
\end{figure}

\end{document}